\documentstyle[preprint,aps]{revtex}
\tighten
\draft
\widetext
\input epsf
\preprint{HUTP-98/A020, NUB 3176}
\bigskip
\bigskip
\begin{document}
\title{On Large $N$ Gauge Theories from Orientifolds}
\medskip
\author{Zurab Kakushadze\footnote{E-mail: 
zurab@string.harvard.edu}}
\bigskip
\address{Lyman Laboratory of Physics, Harvard University, Cambridge, 
MA 02138\\
and\\
Department of Physics, Northeastern University, Boston, MA 02115}
\date{April 28, 1998}
\bigskip
\medskip
\maketitle

\begin{abstract}
{}We consider four dimensional ${\cal N}=1$ supersymmetric gauge theories
obtained via orientifolds of Type IIB on Abelian ${\bf C}^3/\Gamma$ orbifolds.
We construct all such theories that have well defined world-sheet expansion.
The number of such orientifolds is rather limited. We explain this fact in the context
of recent developments in four dimensional Type IIB orientifolds. In particular, we
elaborate these issues in some examples of theories where world-sheet description
is inadequate due to non-perturbative (from the orientifold viewpoint) states
arising from D-branes wrapping (collapsed) 2-cycles in the orbifold. We find complete
agreement with the corresponding statements recently discussed in the context of Type I
compactifications on toroidal orbifolds. This provides a non-trivial check for correctness 
of the corresponding conclusions. We also find non-trivial agreement with
various field theory expectations, and point out their origin in string language.
The orientifold gauge theories that do possess 
well defined world-sheet description have the property that in the large $N$ limit 
computation of any $M$-point correlation function in these theories reduces to
the corresponding computation in the parent ${\cal N}=4$ oriented theory. 
\end{abstract}

\section{Introduction}

{}Recently, motivated by developments in the AdS/CFT correspondence (see, {\em e.g.}, \cite{Kleb,Gubs,Mald,Poly,Oogu,Witt}),
a set of conjectures were proposed \cite{KaSi,LNV} which state that certain gauge theories with
${\cal N}=0,1$ supersymmetries are (super)conformal. These gauge theories are constructed by 
starting from a $U(N)$ gauge theory with ${\cal N}=4$ space-time supersymmetry in four dimensions, and orbifolding by a finite discrete subgroup $\Gamma$ of the $R$-symmetry group
$Spin(6)$ \cite{LNV}. These conjectures were shown at one-loop level for ${\cal N}=0$ theories
\cite{KaSi,LNV}, and to two loops for ${\cal N}=1$ theories using ordinary field theory techniques \cite{LNV}. 

{}In the subsequent development \cite{BKV} these conjectures were shown to be correct to
all loop orders in the large $N$ limit of `t Hooft \cite{thooft}.
The key observation in \cite{BKV} is the 
following. The above gauge theories can be obtained in the $\alpha^\prime\rightarrow 0$ limit
of Type IIB with $N$ parallel D3-branes with the space transverse to the D-branes being ${\bf R}^6/\Gamma$. The gauge theory living in the world-volume of the D3-branes arises as a low energy effective field theory of {\em oriented} open strings that start and end on the D-branes.
The `t Hooft's large $N$ limit then corresponds to taking the limit $N\rightarrow \infty$ with
$\lambda=N\lambda_s$ fixed, where $\lambda_s$ is the Type IIB string coupling. In this context
a world-sheet with $g$ handles (corresponding to closed string loops) and $b$ boundaries (corresponding to D-branes) is weighted with
\begin{equation}\label{thoo}
 (N\lambda_s)^b  \lambda^{2g-2}_s=\lambda^{2g-2+b} N^{-2g+2}~.
\end{equation}  
With the identification $\lambda_s=g_{YM}^2$ we arrive at precisely the large $N$ expansion in the sense of `t Hooft (provided that the effective coupling $\lambda$ is fixed at a weak coupling value)\footnote{A similar expansion was discussed by Witten \cite{CS} for the case of three
dimensional Chern-Simons gauge theory where the boundaries of the string world-sheet are
``topological'' D-branes.}.

{}In \cite{BKV} the above idea was applied to prove that four dimensional 
gauge theories (including the cases with no supersymmetry) considered in
\cite{KaSi,LNV}\footnote{For other related works, see, {\em e.g.}, \cite{Iban,HSU}.} 
are conformal to all orders in perturbation theory in the large $N$ 
limit. The ultraviolet finiteness of string theory (that is, one-loop tadpole cancellation 
conditions) was shown to imply that the resulting (non-Abelian) gauge theories
where conformal in the large $N$ limit (in all loop orders). Moreover, in \cite{BKV}
it was also proven that computation of any correlation function in these theories 
in the large $N$ limit reduces to the corresponding computation in the parent 
${\cal N}=4$ supersymmetric gauge theory.

{}The work in \cite{BKV} was generalized in \cite{zura} where 
the setup was Type IIB string theory with D3-branes as well as orientifold planes
imbedded in the orbifolded space-time. (In certain cases string consistency also requires presence of D7-branes.) This corresponds to Type IIB orientifolds. Introducing orientifold planes
is necessary to obtain $SO$ and $Sp$ gauge groups (without orientifold planes the gauge group
is always unitary), and also allows for additional variety in possible matter content. The presence of orientifold planes changes the possible topologies of the world-sheet.
Now we can have a world-sheet with $b$ boundaries, 
$c$ cross-caps (corresponding to orientifold planes), and $g$ handles. Such a 
world-sheet is weighted with     
\begin{equation}\label{thoo1}
 (N\lambda_s)^b \lambda_s^c \lambda^{2g-2}_s=\lambda^{2g-2+b+c} N^{-c-2g+2}~.
\end{equation}
Note that addition of a cross-cap results in a diagram suppressed by an additional 
power of $N$, so that in the large $N$ limit the cross-cap contributions are subleading.
In fact, in \cite{zura} it was shown that for string vacua which are perturbatively consistent 
(that is, the tadpoles cancel) calculations of correlation functions 
in ${\cal N}<4$ gauge theories
reduce to the corresponding calculations in the parent ${\cal N}=4$
{\em oriented} theory. This holds not only for finite (in the large $N$ limit) gauge theories
but also for the gauge theories which are not conformal. (In the latter case the gauge coupling running was shown to be suppressed in the large $N$ limit.) Here we note that the power of string perturbation techniques is an invaluable tool in proving such statements\footnote{In \cite{BJ}
the proofs of \cite{BKV} were cast into the field theory language. This is straightforward to do
once the string expansion is identified with `t Hooft's large $N$ expansion
as in \cite{BKV} in the case
of unitary gauge theories (that is, in absence of orientifold planes). However, 
string theory gives much more insight than the field theory approach ({\em e.g.}, it explains 
why the orbifold action on Chan-Paton matrices must be in an $n$-fold copy of the regular representation of the orbifold group). In the cases with 
orientifold planes string theory gives an insight for why the number of consistent theories of
this type is so limited, whereas within the field theory this fact seems to be a bit mysterious.}.

{}One distinguishing feature of large $N$ gauge theories obtained via
orientifolds is that the number
of possibilities which possess well defined world-sheet expansion (that is, are perturbative
from the orientifold viewpoint) is rather limited. This might appear surprising at the first sight, since naively one might expect that any choice of the orbifold group $\Gamma$ should lead to a consistent world-sheet theory. As was recently discussed at length in \cite{KST}, in most cases
orientifolds contain non-perturbative sectors (arising from D-branes wrapping collapsed two-cycles at orbifold singularities) which have no world-sheet description. The subject of this note is to elaborate these issues in the case of large $N$ gauge theories obtained as Type IIB
orientifolds. On the other hand, construction of such theories with well defined world-sheet description provides a very non-trivial check for correctness of the conclusions of \cite{KST}
concerning perturbative consistency of various compact orientifold models discussed in the
literature. In particular, our results here are in complete agreement with the conclusions of \cite{KST}.
In this sense we benefit both ways by considering such a setup: ({\em i}) we gain insight into the structure of (large $N$) gauge theories; ({\em ii}) we obtain an independent 
check for rather non-trivial statements concerning world-sheet consistency of Type IIB orientifolds. In this context, this note can be viewed as a companion paper to {\em both} \cite{zura} and \cite{KST}. Nonetheless, this paper is rather self-contained, and can be read 
independently of \cite{zura} or \cite{KST}.

{}The remainder of this paper is organized as follows. In section \ref{prelim} we review
the discussion in \cite{zura}, and also explain some of the facts related to world-sheet consistency of Type IIB orientifolds in the context of \cite{KST}. In section \ref{N1} we construct
all large $N$ gauge theories from orientifolds of Type IIB on ${\bf C}^3/\Gamma$. In this paper we confine our attention ${\cal N}=1$ supersymmetric cases with {\em Abelian} orbifold groups $\Gamma$. In section \ref{other} we discuss other Abelian orbifold groups for which the world-sheet description is inadequate. We explain in detail  the reasons for this, and compare
string theory results with various field theoretic observations. We find a complete agreement with
the conclusions of \cite{KST}, as well as a nice match with field theory expectations. We illustrate
our discussions with various examples. In section \ref{sum} we summarize the main conclusions of this paper.

\section{Preliminaries}\label{prelim}

{}In this section we review the setup in \cite{zura}. Following the recent developments
in \cite{KST}, we then discuss the issues that arise in
four dimensional ${\cal N}=1$ supersymmetric orientifolds. We will apply these discussions to construction of large $N$ gauge theories from
orientifolds in the next section.

\subsection{Setup}

{}Consider Type IIB string theory on ${\bf C}^3/\Gamma$ where
$\Gamma\subset SU(3)$ so that the resulting theory has ${\cal N}=2$ 
supersymmetry in four dimensions. In the following we will confine our
attention to Abelian orbifold groups $\Gamma=\{g_a\vert a=1,\dots,|\Gamma|\}$
($g_1=1$).
Consider the $\Omega J$ orientifold of this 
theory, where $\Omega$ is the world-sheet parity reversal, and $J$ 
is a ${\bf Z}_2$ element ($J^2=1$) acting on the complex coordinates $z_i$
($i=1,2,3$) on ${\bf C}^3$ as follows: $J z_i=-Jz_i$. The resulting theory has ${\cal N}=1$
supersymmetry in four dimensions. 
 
{}Note that we have an orientifold 3-plane corresponding to the $\Omega J$
element of the orientifold group. If $\Gamma$ has
a ${\bf Z}_2$ subgroup, then we also have an orientifold 7-plane.
If we have an orientifold 7-plane we must 
introduce 8 of the corresponding D7-branes to cancel the R-R charge appropriately.
(The number 8 of D7-branes is required by the corresponding tadpole cancellation
conditions.) Note, however, that the number of D3-branes is not constrained (for the corresponding untwisted tadpoles automatically vanish in the non-compact case).

{}We need to specify the action of $\Gamma$ on the Chan-Paton factors
corresponding to the D3- and D7-branes.  
These are given by Chan-Paton matrices which we collectively refer to
as $\gamma^\mu_a$, where the superscript $\mu$ refers to the corresponding
D3- or D7-branes. Note that ${\mbox{Tr}}(\gamma^\mu_1)=n^\mu$ where 
$n^\mu$ is the number of D-branes labelled by $\mu$. 

{}At one-loop level there are three different sources for massless tadpoles:
the Klein bottle, annulus, and M{\"o}bius strip amplitudes. The factorization property of string theory implies that the tadpole cancellation conditions read (see, {\em e.g.}, \cite{zura}
for a more detailed discussion):
\begin{equation}\label{BC}
 B_a+\sum_\mu C^\mu_a {\mbox{Tr}}(\gamma^\mu_a)=0~.
\end{equation}
Here $B_a$ and $C^\mu_a$ are (model dependent) numerical coefficients of order 1. 

{}In the world-volume of D3-branes there lives a four dimensional ${\cal N}=1$ supersymmetric gauge theory (which is obtained in the low energy, that is, $\alpha^\prime\rightarrow 0$ limit). Since the number of D3-branes is unconstrained, we can consider the large
$N$ limit of this gauge theory. In \cite{zura} (generalizing the work in \cite{BKV}) it was shown that, if for a given
choice of the orbifold group $\Gamma$ the world-sheet description for the orientifold is adequate, 
then in the large $N$ limit (with $N\lambda_s$ fixed, where $\lambda_s$ is the Type IIB string coupling) computation of any correlation function in this gauge theory is reduced to the corresponding computation in the parent ${\cal N}=4$ supersymmetric {\em oriented} gauge theory before orbifolding and orientifolding. In particular, the $\beta$-function coefficients grow
as
\begin{equation}\label{beta}
 b_s=O(N^s)~,~~~s=0,1,\dots~,
\end{equation}
as opposed to $b_s=O(N^{s+1})$ (as in, say, pure $SU(N)$ gauge theory). 
This implies that the running of the gauge couplings in the large $N$ limit is suppressed. Moreover, it was also pointed out that if the tadpole cancellation conditions (\ref{BC}) imply that
\begin{equation}\label{Klein}
 {\mbox {Tr}}(\gamma^\mu_a)=0~\forall a\not=1
\end{equation}
(that is, $B_a=0$ $\forall a\not=1$), then the the one-loop $\beta$-function coefficients $b_0$
for non-Abelian gauge theories living in world-volumes of the D3-branes vanish.

{}The reason for the above properties of gauge theories coming from the D3-branes
is not difficult to understand. The information about the fact that the orbifold theory has reduced supersymmetry is encoded in the action of the twists $g_a$ ($a\not=1$) which are accompanied by the corresponding Chan-Paton matrices $\gamma^\mu_a$, and also in the presence of cross-caps
on the world-sheet which correspond to the action of the orientifold projection $\Omega$. In the large $N$ limit contributions of these twists and cross-caps into the correlation functions are subleading as they are always of order 1 (recall, for instance, that all $B_a$ and 
$C^\mu_a$ are of order 1). This implies that the leading large $N$ 
contribution comes from trivial (that is, untwisted) boundary conditions associated with D3-brane
boundaries on the world-sheet. In particular, all contributions corresponding to 
D7-branes (whose number is of order 1), cross-caps, handles (which correspond to closed
string loops) and twisted D3-branes are subleading in the large $N$ limit, that is, when the
number of D3-branes is large. The remaining contributions (up to overall numerical factors)
are the same as in the parent ${\cal N}=4$ {\em oriented} gauge theory (before orbifolding and orientifolding). We refer the reader to \cite{BKV} and \cite{zura} for more detail.

\subsection{Perturbative Orientifolds}

{}The arguments of \cite{zura} that imply the above properties of D3-brane gauge theories
are intrinsically perturbative. In particular, a consistent world-sheet expansion is crucial for their
validity. It is therefore important to understand the conditions for the perturbative orientifold
description to be adequate. 

{}Naively, one might expect that any choice of the orbifold group $\Gamma\subset Spin(6)$
(note that $Spin(6)$ is the $R$-symmetry group of ${\cal N}=4$ gauge theory)
should lead to an orientifold with well defined world-sheet expansion in terms of boundaries
(corresponding to D-branes), cross-caps (corresponding to orientifold planes) and handles
(corresponding to closed string loops). This is, however, not the case \cite{KST}. In fact, the number of choices of $\Gamma$ for which such a world-sheet expansion is adequate is rather
constrained. Before we consider these cases in detail, let us discuss the origins of these constraints.  
  
{}Thus, consider the $\Omega J$ orientifold of Type IIB on ${\bf C}^3/\Gamma$. The orientifold
group is given by ${\cal O}=\{g_a,\Omega Jg_a\vert a=1,\dots, |\Gamma|\}$. The sectors labeled by $g_a$ correspond to the unoriented closed twisted plus untwisted sectors. The sectors labeled by $\Omega Jg_a$ with $(Jg_a)^2=1$ correspond to open strings stretched between D-branes. (In particular, if the set of points fixed under $Jg_a$ has dimension 0
then these are D3-branes. If this set has real dimension 4, then these are D7-branes.) However,
the sectors labeled by $\Omega Jg_a$ with $(Jg_a)^2\not=1$ do not have an interpretation in
terms of open strings starting and ending on perturbative D-branes ({\em i.e.}, they do not have an interpretation in terms of open strings with purely Dirichlet or Neumann boundary conditions
in all directions) \cite{KST}. Instead, if viewed as open strings they would have mixed (that is, neither Dirichlet nor Neumann) boundary conditions. These states do not have world-sheet
description. They can be viewed as arising from D-branes wrapping (collapsed) two-cycles in the
orbifold \cite{KST}. These states are clearly non-perturbative from the orientifold viewpoint.

{}This difficulty is a generic feature in most of the orientifolds of Type IIB compactified on toroidal
orbifolds, as well as the corresponding non-compact cases such as the $\Omega J$ orientifolds of Type IIB on ${\bf C}^3/\Gamma$. However, there is a (rather limited) class of cases where the would-be non-perturbative
states are massive (and decouple in the low energy effective field theory) if we consider
compactifications on blown up orbifolds \cite{KST}. In fact, these blow-ups are forced by
the orientifold consistency.
The point is that the orientifold projection $\Omega$ must be chosen to be the same 
as in the case of Type IIB on a smooth Calabi-Yau three-fold. 
The reasons why this choice of the orientifold
projection is forced have been recently discussed at length in \cite{KST}. In particular, we do not
have an option of choosing the orientifold projection analogous to that in the six dimensional models of \cite{GJ}. On the other hand, the above $\Omega$ orientifold projection is
not a symmetry of Type IIB on ${\bf C}^3/\Gamma$ at the orbifold conformal field theory point \cite{KST}.
The reason for this is that $\Omega$ correctly reverses the world-sheet orientation of world-sheet bosonic and fermionic oscillators and left- and right-moving momenta, but fails
to do the same with the {\em twisted} ground states. (Such a reversal would involve mapping  
the $g_a$ twisted ground states to the $g^{-1}_a$ twisted ground states.
In \cite{KST} such an orientation reversal was shown to be inconsistent.) This difficulty is circumvented by noting that the orientifold projection $\Omega$ is consistent for
{\em smooth} Calabi-Yau three-folds, and, in particular, for a blown up version of the ${\bf C}^3/\Gamma$
orbifold. Thus, once the appropriate blow-ups are performed, the orientifold procedure is well
defined.

{}In some cases the blow-ups result in decoupling of the would-be massless non-perturbative
states, which is due to the presence of an appropriate superpotential (that couples the blow-up
modes to the non-perturbative states). This feature, however, is not generic and is only present
in a handful of cases. This was shown to be the case in $\Omega$ orientifolds of Type IIB on
$T^6/\Gamma$ with $\Gamma\approx{\bf Z}_3$, ${\bf Z}_7$, ${\bf Z}_3\otimes 
{\bf Z}_3$ in \cite{ZK,KS1,KS2}. These orientifolds correspond to Type I compactifications on
the corresponding orbifolds, which in turn have perturbative heterotic duals (the corresponding heterotic compactifications are perturbative as there are no D5-branes (which would map to heterotic NS5-branes) in these models). The non-perturbative (from the orientifold viewpoint) states were
shown to correspond to twisted sector states on the heterotic side. The perturbative superpotentials for these states can be readily computed, and are precisely such that after the
appropriate blow-ups (those needed for the orientifold consistency) the twisted sector states decouple.

{}These arguments were generalized in \cite{KST} to the ${\bf Z}_2\otimes {\bf Z}_3$
model of \cite{KS2} and the ${\bf Z}_2\otimes {\bf Z}_2\otimes {\bf Z}_3$ model of \cite{zk}.
In all the other cases (except for the ${\bf Z}_2\otimes {\bf Z}_2$ model of \cite{BL} which is
obviously perturbative from the above viewpoint) it was argued in \cite{KST} that non-perturbative states do not decouple. Various checks of these statements were performed in
\cite{KST} using the web of dualities between Type IIB orientifolds, F-theory, and Type I and heterotic compactifications on orbifolds. These statements, however, only depend on local properties
of D-branes and orientifold planes near orbifold singularities and should persist in non-compact
cases such as the $\Omega J$ orientifolds of Type IIB on ${\bf C}^3/\Gamma$. By studying 
the latter we can therefore perform an independent check of the conclusions of \cite{KST}.
On the other hand, this would also provide an understanding of large $N$ gauge theories from
orientifolds (and, in particular, why their number is so limited). The next two sections are devoted
to precisely this subject. 

\section{${\cal N}=1$ Gauge Theories}\label{N1} 

{}In this section we construct ${\cal N}=1$ orientifolds of Type IIB on
${\bf C}^3/\Gamma$ with $\Gamma\approx{\bf Z}_3$, ${\bf Z}_7$, ${\bf Z}_3\otimes 
{\bf Z}_3$, ${\bf Z}_2\otimes {\bf Z}_2$, ${\bf Z}_2\otimes {\bf Z}_2\otimes 
{\bf Z}_3$ and ${\bf Z}_2\otimes {\bf Z}_3$. (These are understood to be subgroups
of $SU(3)$.) We first discuss models without D7-branes (that is, the ${\bf Z}_3$, ${\bf Z}_7$ and
${\bf Z}_3\otimes {\bf Z}_3$ cases), and then consider the other three models with
D7-branes. 

\subsection{Theories without D7-branes}

{}Consider the $\Omega J$ orientifold of Type IIB on ${\bf C}^3/\Gamma$ where the
orbifold group $\Gamma=\{g^k\vert k=0,\dots,M-1\}\approx {\bf Z}_M$ ($M$ is odd)
is a subgroup of $SU(3)$ (but not of $SU(2)$). The action of $\Gamma$ on the complex 
coordinates $z_i$ ($i=1,2,3$) on ${\bf C}^3$ is given by $gz_i=\omega^{\ell_i} z_i$, 
where $\omega=\exp(2\pi i/M)$, $\ell_i\not=0$, and 
without loss of generality we can take $\ell_1=1$, $\ell_2=p$, $\ell_3=M-p-1$ with
$p\in\{1,\dots,M-2\}$.

{}In these orientifolds we have only D3-branes whose number is arbitrary. The twisted 
tadpole cancellation conditions for these orientifolds are isomorphic (upon interchanging
the corresponding D3- and D9-brane Chan-Paton matrices) 
to those for the $\Omega$ orientifolds of Type IIB on ${\bf C}^3/\Gamma$. The latter  
tadpole cancellation conditions were derived in \cite{KS1}. Applying those results to
the cases under consideration we have the following twisted tadpole cancellation 
conditions ($k=1,\dots,N-1$):
\begin{equation}\label{tadpoles1}
 {\mbox{Tr}}(\gamma_{2k,3})= -4\eta\prod_{i=1}^3 (1+\omega^{k\ell_i})~.
\end{equation}    
Here $\eta=-1$ if the $\Omega$ projection is of the $SO$ type, and $\eta=+1$ if it is of the
$Sp$ type.

\begin{center}
 {\em The} ${\bf Z}_3$ {\em Model}
\end{center}

{}First consider the case where $M=3$. This model is a ``T-dual'' (in the non-compact limit)
of the model studied in \cite{Sagnotti}\footnote{Also see, {\em e.g.}, \cite{LPT}.}. 
The solution to the twisted tadpole cancellation
conditions reads ($N=(n_3-4\eta)/3$):
\begin{eqnarray}
 \gamma_{1,3}={\mbox{diag}}(
 \exp(2\pi i/3)~(N~{\mbox{times}}),
 \exp(-2\pi i/3)~(N~{\mbox{times}}),
 1~({N+4\eta}~{\mbox{times}}))~.
\end{eqnarray}
The massless spectra (for both choices of $\eta$) of these models are given in
Table \ref{Z3}. The non-Abelian gauge anomaly is cancelled in this model.
Let us denote the fields in the 33 open string sector as follows (see Table \ref{Z3}):
$\Phi_i=3\times ({\bf R}_\eta,{\bf 1})(+2)$ and $Q_i=3\times ({\overline {\bf N}},{\bf N+4\eta})(-1)$.
Here $i(=1,2,3)$ labels the multiplicity of the fields. The superpotential of this model is given
by \cite{Sagnotti,ZK}:
\begin{equation}
 {\cal W}=\epsilon_{ijk}\Phi_i Q_j Q_k~.
\end{equation}
Here and in the following the summation over repeated indices (in expressions for
superpotentials) is understood. Also, we will always suppress the actual values of the Yukawa
coupling and only display the {\em non-vanishing} terms.

\begin{center}
 {\em The} ${\bf Z}_7$ {\em Model}
\end{center}

{}Next, let us consider the case where $M=7$. This model is a ``T-dual'' (in the non-compact limit) of the model studied in \cite{KS1}.
The solution to the twisted tadpole cancellation
conditions reads ($N=(n_3+4\eta)/7$):
\begin{eqnarray}
 \gamma_{1,3}={\mbox{diag}}(&&
 \exp(2\pi i/7)~(N~{\mbox{times}}),
 \exp(-2\pi i/7)~(N~{\mbox{times}}),\nonumber\\
 &&\exp(4\pi i/7)~(N~{\mbox{times}}),
 \exp(-4\pi i/7)~(N~{\mbox{times}}),\nonumber\\
 &&\exp(6\pi i/7)~(N~{\mbox{times}}),
 \exp(-6\pi i/7)~(N~{\mbox{times}}),1~({N-4\eta}~{\mbox{times}}))~.
\end{eqnarray}
The massless spectra (for both choices of $\eta$) of these models are given in
Table \ref{Z3}. The non-Abelian gauge anomaly is cancelled in this model.
Let us denote the fields in the 33 open string sector as follows (see Table \ref{Z3}):
$\Phi_1= ({\bf 1},{\bf 1},{\bf R}_\eta,{\bf 1})(0,0,+2)$,
$P_1= ({\bf N},{\bf 1},{\bf 1},{\bf N-4\eta})(+1,0,0)$,
$R_1=({\overline {\bf N}},{\bf N},{\bf 1},{\bf 1})(-1,+1,0)$, 
$Q_1=({\bf 1},{\overline {\bf N}},{\overline {\bf N}},{\bf 1})(0,-1,-1)$, plus cyclic permutations
of the $U(N)\otimes U(N)\otimes U(N)$ irreps which define $\Phi_i, P_i,R_i, Q_i$ with $i=2,3$.
The superpotential of this model is given
by \cite{KS2}:
\begin{equation}
 {\cal W}=\epsilon_{ijk} P_i P_j Q_k+\epsilon_{ijk} Q_i R_j \Phi_k +
 \epsilon_{ijk} R_i R_j R_k~.
\end{equation}

\begin{center}
 {\em The} ${\bf Z}_3\otimes {\bf Z}_3$ {\em Model}
\end{center}

{}Let us now consider the case with $\Gamma\approx{\bf Z}_3\otimes {\bf Z}_3$. 
This model is a ``T-dual'' (in the non-compact limit)
of the model studied in \cite{KS2}.
Let $g_1$ and $g_2$ be the generators of the two ${\bf Z}_3$ subgroups. Their action
on the complex coordinates $z_i$ is given by: $g_1z_1=\omega z_1$, 
$g_1z_2=\omega^{-1} z_2$, $g_1z_3= z_3$, $g_2z_1= z_1$, 
$g_2z_2=\omega z_2$, $g_2z_3=\omega^{-1} z_3$, where $\omega=\exp(2\pi i/3)$.  
The twisted tadpole cancellation conditions read \cite{KS2,zura}\footnote{The tadpole cancellation conditions for ${\mbox{Tr}}(\gamma_{g_1,3})$ and 
${\mbox{Tr}}(\gamma_{g_2,3})$ follow from those of \cite{GJ} (where
the corresponding tadpole cancellation conditions were derived for a system of D9- and/or
D5-branes) upon compactifying on $T^2$, T-dualizing, and taking the dimensions of the dual torus ${\widetilde T}^2$ to infinity. The orientifold $p$-planes ($p=5,9$) split into 4 orientifold 
$(p-2)$-planes upon T-dualizing. These are located at four fixed points of 
${\widetilde T}^2/{\bf Z}_2$. However, after taking the dimensions of 
${\widetilde T}^2$ to infinity ({\em i.e.}, when considering ${\bf C}/{\bf Z}_2$ instead of
${\widetilde T}^2/{\bf Z}_2$), only one fixed point (at the origin) remains. This results in 
reduction of all the tadpoles by a factor of 4. In particular, the number of D7-branes is 8 as opposed to 32 in the case of D9-branes.}
\begin{eqnarray}
 {\mbox{Tr}}(\gamma_{g_1,3})= {\mbox{Tr}}(\gamma_{g_2,3})=
 {\mbox{Tr}}(\gamma_{g_1g_2,3})=-2\eta~,\\
 {\mbox{Tr}}(\gamma_{g_1g^2_2,3})={\mbox{Tr}}(\gamma_{g^2_1g_2,3})=4\eta~.
\end{eqnarray} 
The solution to the twisted tadpole cancellation
conditions reads ($N=(n_3-4\eta)/9$):
\begin{eqnarray}
 \gamma_{g_1,3}={\mbox{diag}}(&&
 \exp(2\pi i/3)~(3N+2\eta~{\mbox{times}}),
 \exp(-2\pi i/3)~(3N+2\eta~{\mbox{times}}),\nonumber\\
 &&1~(3N~{\mbox{times}}))~,\\
\gamma_{g_2,3}={\mbox{diag}}(&&
 \exp(2\pi i/3)~(N+2\eta~{\mbox{times}}),
 \exp(-2\pi i/3)~(N~{\mbox{times}}),1~(N~{\mbox{times}}),\nonumber\\
 &&
 \exp(-2\pi i/3)~(N+2\eta~{\mbox{times}}),
 \exp(2\pi i/3)~(N~{\mbox{times}}),1~(N~{\mbox{times}}),\nonumber\\
 &&
 \exp(2\pi i/3)~(N~{\mbox{times}}),
 \exp(-2\pi i/3)~(N~{\mbox{times}}),1~(N~{\mbox{times}}))~.
\end{eqnarray}
The massless spectra (for both choices of $\eta$) of these models are given in
Table \ref{Z3}. The non-Abelian gauge anomaly is cancelled in this model.
Let us denote the fields in the 33 open string sector as follows (see Table \ref{Z3}):
$\chi_1= ({\bf 1}, {\overline {\bf N}},{\bf 1},{\bf 1},{\bf N})(0,-1,0,0)$,
$P_1=({\bf 1},{\overline {\bf N}},{\bf 1}, {\bf N+2\eta},{\bf 1})(0,-1,0,+1)$, 
$Q_1=({\bf 1},{\bf 1}, {\overline {\bf N}},{\overline {\bf N+2\eta}},{\bf 1})(0,0,-1,-1)$, 
$R_1=({\bf 1},{\bf N},{\bf N},{\bf 1},{\bf 1})(0,+1,+1,0)$,
plus cyclic permutations
of the $U(N)\otimes U(N)\otimes U(N)$ irreps which define $\chi_i, P_i,R_i, Q_i$ with $i=2,3$.
The superpotential of this model is given by:
\begin{equation}
 {\cal W}=\epsilon_{ijk} \chi_i \chi_j R_k+\epsilon_{ijk} P_i Q_j R_k~.
\end{equation}

\subsection{Theories with D7-branes}

{}Now let us consider the $\Omega J$ orientifolds of Type IIB on ${\bf C}^3/\Gamma$,
where $\Gamma\approx{\bf Z}_2\otimes {\bf Z}_2$, ${\bf Z}_2\otimes {\bf Z}_2
\otimes {\bf Z}_3$ and ${\bf Z}_2\otimes {\bf Z}_3$. All of these models contain D7-branes
as $\exists {\bf Z}_2\subset \Gamma$.

\begin{center}
 {\em The} ${\bf Z}_2\otimes {\bf Z}_2$ {\em Model}
\end{center}

{}We start from Type IIB string theory on 
${\bf C}^3/\Gamma$, where $\Gamma=\{1,R_1,R_2,R_3\}\approx {\bf Z}_2
\otimes {\bf Z}_2$ ($R_i R_j=R_k$, $i\not=j\not=k\not=i$) is the
orbifold group whose action on the complex coordinates $z_i$
is given by $R_i z_j=-(-1)^{\delta_{ij}} z_j$. 
Next, we consider an orientifold of this theory where the 
orientifold action is given by $\Omega J$.

{}This model is a ``T-dual'' (in the non-compact limit) of the model studied in \cite{BL}.
The untwisted tadpole cancellation conditions require presence of three sets of 
D7-branes with 8 D7-branes in each set. Thus, the locations of D$7_i$-branes are 
given by points in the $z_i$ complex plane. The number of D3-branes is unconstrained.
The twisted
tadpole cancellation conditions imply that the corresponding Chan-Paton matrices
$\gamma_{R_i,3}$ and $\gamma_{R_i,7_j}$ are traceless:
\begin{equation}
 {\mbox{Tr}}(\gamma_{R_i,3})={\mbox{Tr}}(\gamma_{R_i,7_j})=0~.
\end{equation} 
A choice\footnote{This choice is unique up to equivalent representations.
The uniqueness of this choice was argued in \cite{BL} from the orientifold viewpoint, and was recently shown in \cite{KST} from F-theory \cite{vafa} considerations.} 
consistent
with requirements that the Chan-Paton matrices form a (projective) representation
of (the double cover) of $\Gamma$ is given by ($N=n_3/2$)
\begin{eqnarray}
 \gamma_{R_i,3}=i\sigma_i\otimes {\bf I}_N~,
\end{eqnarray} 
where $\sigma_i$ are Pauli matrices, and ${\bf I}_N$ is an $N\times N$ identity matrix.
(The action on the D$7_i$ Chan-Paton charges
is similar.) The spectrum of this model is given in Table \ref{Z2}.
Let
$\Phi_i$, $\Phi^i_j$, $Q^i$ and $Q^{ij}$ be the matter fields in the 33, $7_i 7_i$,
$37_i$ and $7_i7_j$ open string sectors, respectively. The subscript in 
$\Phi_i$ and $\Phi^i_j$ labels three different chiral superfields (see Table 
\ref{Z2}) in the 33 and $7_i 7_i$ sectors. The superpotential of this model
is given by \cite{zura}
\begin{equation}
 {\cal W}=\epsilon_{ijk}\Phi_i\Phi_j\Phi_k + 
 \epsilon_{ijk}\Phi^l_i\Phi^l_j\Phi^l_k +
 \epsilon_{ijk}\Phi^i_k Q^{ij} Q^{ij} +
 \Phi_i Q^i Q^i+
 Q^{ij} Q^{jk} Q^{ki} +
 Q^{ij} Q^i Q^j~.
\end{equation}
 
\begin{center}
 {\em The} ${\bf Z}_2\otimes {\bf Z}_2\otimes {\bf Z}_3$ {\em Model}
\end{center}

{}Let us now consider the case with $\Gamma
\approx {\bf Z}_2\otimes {\bf Z}_2\otimes {\bf Z}_3$. Let
$R_i$ be the same as in the previous example, and let $g$ be the generator of   
${\bf Z}_3\subset \Gamma$ which acts as in the ${\bf Z}_3$ model.
This model is a ``T-dual'' (in the non-compact limit) of the model studied in \cite{zk}.
Just as in the previous case, 
the untwisted tadpole cancellation conditions require presence of three sets of 
D7-branes with 8 D7-branes in each set. The twisted
tadpole cancellation conditions read \cite{KS2,zk} ($k=1,2$):
\begin{eqnarray}
 &&{\mbox{Tr}}(\gamma_{R_i,3})={\mbox{Tr}}(\gamma_{R_i,7_j})=0~,\\
 &&{\mbox{Tr}}(\gamma_{g^kR_i,3})={\mbox{Tr}}(\gamma_{g^kR_i,7_j})=0~\\
 &&{\mbox{Tr}}(\gamma_{g^k,3})={\mbox{Tr}}(\gamma_{g^k,7_j})=-4~.
\end{eqnarray} 
The solution to the twisted tadpole cancellation conditions reads ($N=(n_3+4)/6$):
\begin{eqnarray}
 &&\gamma_{g,3}={\mbox{diag}}(\omega {\bf I}_{2N},\omega^2 {\bf I}_{2N},{\bf I}_{2N-4})~,\\
 &&\gamma_{R_i,3}=i\sigma_i\otimes {\bf I}_{3N-2}~.
\end{eqnarray}
(Here $\omega=\exp(2\pi i/3)$.) 
The action on D$7_i$ Chan-Paton charges is similar except that their number is fixed:
$n_{7_i}=8$. The spectrum of this model is given in Table \ref{Z2}. The non-Abelian gauge
anomaly is cancelled in this model. Let
$\Phi_i=3\times ({\bf A},{\bf 1})(+2)_{33}$,
$\chi_i=3\times ({\overline {\bf N}}, {\bf N-2}) (-1)_{33}$, 
$P^i=({\bf N},{\bf 1};{\bf 2}_i)(+1;+1_i)_{37_i}$,
$R^i=({\bf 1},{\bf N-2};{\bf 2}_i)(0,-1_i)_{37_i}$ (see Table \ref{Z2}).
The superpotential of this model
is given by
\begin{equation}
 {\cal W}=\epsilon_{ijk}\Phi_i\chi_j\chi_k + \chi_iP^iR^i~.
\end{equation}

\begin{center}
 {\em The} ${\bf Z}_2\otimes {\bf Z}_3$ {\em Model}
\end{center}

{}Finally, consider the case with $\Gamma\approx{\bf Z}_2\otimes {\bf Z}_3$. Here
the action of the generators $R$ of ${\bf Z}_2\subset \Gamma$ and $g$ of ${\bf Z}_3\subset
\Gamma$ on the complex coordinates $z_i$ is given by: $Rz_1=-z_1$, $Rz_2=-z_2$, $Rz_3=z_3$, $gz_i=\omega z_i$ ($\omega=\exp(2\pi i/3)$). This model is a ``T-dual'' (in the non-compact limit) of the model studied in \cite{KS2}. The untwisted tadpole cancellation 
conditions require presence of {\em one} set of 8 D7-branes (whose locations are given by
points in the $z_3$ complex plane). The twisted tadpole cancellation conditions read
\cite{KS2} ($k=1,2$):
\begin{eqnarray}
 &&{\mbox{Tr}}(\gamma_{R,3})={\mbox{Tr}}(\gamma_{R,7})=0~,\\
 &&{\mbox{Tr}}(\gamma_{g^kR,3})={\mbox{Tr}}(\gamma_{g^kR,7})=0~\\
 &&{\mbox{Tr}}(\gamma_{g^k,3})={\mbox{Tr}}(\gamma_{g^k,7})=-4~.
\end{eqnarray} 
The solution to the twisted tadpole cancellation conditions reads ($N=(n_3+4)/6$):
\begin{eqnarray}
 &&\gamma_{g,3}={\mbox{diag}}(\omega {\bf I}_{2N},\omega^2 {\bf I}_{2N},{\bf I}_{2N-4})~,\\
 &&\gamma_{R,3}={\mbox{diag}}(i,-i)\otimes {\bf I}_{3N-2}~,\\
 &&\gamma_{g,7}={\mbox{diag}}(\omega {\bf I}_{4},\omega^2 {\bf I}_{4})~,\\
 &&\gamma_{R,7}={\mbox{diag}}(i,-i)\otimes {\bf I}_{4}~.
\end{eqnarray}
The spectrum of this model is given in Table \ref{Z2}. The non-Abelian gauge
anomaly is cancelled in this model. Let 
$\Phi_a=2\times ({\bf A}, {\bf 1},
{\bf 1}) (+2,0,0)_{33}$,
${\widetilde \Phi}_a=2\times ({\bf 1},{\overline {\bf A}}, 
{\bf 1}) (0,-2,0)_{33}$, 
$Q_a=2\times ({\overline {\bf N}}, {\bf 1},
{\overline {\bf N-2}}) (-1,0,-1)_{33}$, ${\widetilde Q}_a=2\times ({\bf 1},{\bf N}, 
{\bf N-2}) (0,+1,+1)_{33}$. Next, $P=({\bf N},{\overline {\bf N}},
{\bf 1}) (+1,-1,0)_{33}$, $R= ({\overline {\bf N}}, {\bf 1},
{\bf N-2}) (-1,0,+1)_{33}$, ${\widetilde R}= ( {\bf 1},{\bf N},
{\overline {\bf N-2}}) (0,+1,-1)_{33}$, 
$S= ({\bf N}, {\bf 1},{\bf 1};{\bf 2},{\bf 1})(+1,0,0;+1,0)_{37}$,
${\widetilde S}=({\bf 1},{\overline {\bf N}},{\bf 1};{\bf 1},{\bf 2})(0,-1,0;0,-1)_{37}$. 
We will also use
$T=({\bf 1},{\bf 1},{\bf N-2};{\bf 1},{\bf 2})(0,0,+1;0,+1)_{37}$,
${\widetilde T}=({\bf 1},{\bf 1},{\overline {\bf N-2}};{\bf 2},{\bf 1})(0,0,-1;-1,0)_{37}$,
$U=({\bf 2},{\bf 2})(+1,-1)_{77}$ (see Table \ref{Z2}).
Here $a=1,2$ labels the multiplicity of states in the
33 open string sector. The superpotential of this model is given by:
\begin{eqnarray}
 {\cal W}=&&\Phi_1 Q_2R+\Phi_2 Q_1 R+{\widetilde\Phi_1} {\widetilde Q_2}{\widetilde R}+
 {\widetilde \Phi}_2 {\widetilde Q}_1 {\widetilde R}+\nonumber\\
 &&Q_1{\widetilde Q_2}P+Q_2{\widetilde Q_1}P+T{\widetilde T}U+
 T{\widetilde S}{\widetilde R}+{\widetilde T} S R~.
\end{eqnarray}

\subsection{Comments}

{}Note that the one-loop $\beta$-function of the D3-brane gauge theory vanishes
in the ${\bf Z}_2\otimes {\bf Z}_2$ model. In the other five models 
the one-loop $\beta$-functions of the D3-brane gauge theories (more precisely,
their non-Abelian parts) do not vanish. This is in accord with the discussion in \cite{zura}.
Namely, the one-loop $\beta$-function is not expected to vanish unless all twisted
Chan-Paton matrices are traceless\footnote{This is expected to hold generically. Nonetheless,
``accidental'' cancellations can occur as in the ${\cal N}=2$ examples of \cite{zura}. In particular, those ``accidental'' cancellations were explained in \cite{zura} based on the results
of \cite{NS}.}.

{}Nonetheless, as was shown in \cite{zura}, in the large $N$ limit (where the effective
gauge coupling $\lambda_s N$ is kept at a finite fixed value, $\lambda_s$ being the Type IIB
string coupling), computation of any correlation function in these gauge theories is reduced
to the corresponding computation in the parent ${\cal N}=4$ {\em oriented} gauge theory. In particular,
the running of the gauge coupling is suppressed in the large $N$ limit. 

{}Here the following remark is in order. In all of the above models, except for the
${\bf Z}_2\otimes {\bf Z}_2$ case, there is present at least one anomalous $U(1)$ \cite{anom}
in the
massless open string spectrum. (More precisely, there are as many anomalous $U(1)$'s as
different types of D-branes. For instance, in the ${\bf Z}_2\otimes {\bf Z}_2\otimes {\bf Z}_3$
model there are 4 anomalous $U(1)$'s: one coming from the 33 open string sector, and the other three coming from the $7_i 7_i$ open string sectors.) Presence of these anomalous $U(1)$'s
implies that there are corresponding Fayet-Iliopoulos D-terms which must be cancelled 
via a generalized Green-Schwarz mechanism \cite{GS}. 
The fields responsible for breaking these $U(1)$'s are some of the closed string sector singlets
(which transform
non-trivially under the anomalous $U(1)$ gauge transformations) corresponding to the
orbifold blow-up modes \cite{Sagnotti,ZK,KST}. On the other hand, as we already 
discussed, in all the cases except for the ${\bf Z}_2\otimes {\bf Z}_2$ model the orbifold
singularities must be blown-up for the orientifold action to be consistent. In this process all
the non-perturbative (from the orientifold viewpoint) states naively expected in the massless
spectrum acquire masses via appropriate superpotentials \cite{ZK,KS1,KS2,KST}. Thus, the
masses of these states are of order $M_X$, that is, the anomalous $U(1)$ scale. This scale
is given by $M_X\sim M_s$ (up to a numerical constant independent of $M_s$ and $\lambda_s$), where $M_s=1/\sqrt{\alpha^\prime}$ is the string scale. Since in the field theory limit we are taking $M_s\rightarrow\infty$, all of these states decouple from the massless modes,
and therefore have no effect on the validity of the perturbative arguments in \cite{zura} which lead
to the statements about the large $N$ behavior of the above gauge theories.

{}Note that in the cases where non-perturbative (from the orientifold viewpoint) states do not
decouple after blow-ups, the arguments of \cite{zura} would not be  be valid. In fact, in such cases we would not expect
those arguments to hold to begin with since the perturbative (that is, world-sheet) description of
the orientifold leads to anomalies as we discuss in the next section. In particular, in those cases
there always remain some uncanceled tadpoles.     

{}Here we should mention that the tadpole cancellation 
is necessary not only for ultraviolet finiteness of the corresponding theories but also
for non-Abelian gauge anomaly cancellation. For illustrative purposes, to see what 
can go wrong if we relax these conditions, let us consider the above ${\bf Z}_3$ example
with the following choice for the twisted Chan-Paton matrices:
\begin{eqnarray}
 \gamma_{1,3}={\mbox{diag}}(
 \exp(2\pi i/3)~(N~{\mbox{times}}),
 \exp(-2\pi i/3)~(N~{\mbox{times}}),
 1~(N^\prime~{\mbox{times}}))~.
\end{eqnarray}  
The gauge group in the 33 open string sector is $U(N)\otimes G_\eta (N^\prime)$, and
the chiral matter is given by $3({\bf R}_\eta,{\bf 1})$ and $3({\overline {\bf N}}, {\bf N}^\prime)$
(see Table \ref{Z3} for notation). Note that ${\bf R}_\eta$ of $SU(N)$ contributes
as much as $N+4\eta$ chiral superfields in ${\bf N}$ of $SU(N)$ into the non-Abelian gauge
anomaly. Thus, the non-Abelian gauge anomaly cancellation implies that 
$N^\prime=N+4\eta$. This is precisely the solution to the tadpole cancellation condition 
(\ref{tadpoles1}).

\section{Other Cases}\label{other}

{}In this section we discuss $\Omega J$ orientifolds of Type IIB on ${\bf C}^3/\Gamma$, where
the orbifold group $\Gamma$ (which is assumed to be an {\em Abelian} subgroup of $SU(3)$) 
is {\em not} among the six cases considered in the previous section. In particular, we show that for such orbifolds it is impossible to cancel all tadpoles (corresponding to the ``naive'' 
world-sheet approach to the orientifold). We then explain this in the light of
the discussions in \cite{KST}.

\subsection{Cases with D7-branes}

{}Let us first consider Abelian orbifolds such that $\Gamma$ contains a ${\bf Z}_2$ subgroup.
This implies that we have to introduce D7-branes. It is not difficult to check that the ``naive'' tadpole cancellation conditions (derived in the world-sheet approach) do not have a solution
(unless $\Gamma$ is one of the three groups, namely, ${\bf Z}_2\otimes {\bf Z}_2$, 
${\bf Z}_2\otimes {\bf Z}_2\otimes {\bf Z}_3$ or ${\bf Z}_2\otimes {\bf Z}_3\approx{\bf Z}_6$
discussed in the previous section).
Instead of being most general here, for illustrative purposes we will consider a particular 
example. Let $\Gamma\approx {\bf Z}_6^\prime$ (we are using prime here to avoid confusion
with the previously studied case of $\Gamma\approx {\bf Z}_2\otimes {\bf Z}_3\approx{\bf Z}_6$) where the generators $g$ and $R$ of the ${\bf Z}_3\subset \Gamma$ and 
${\bf Z}_2\subset \Gamma$ subgroups have the following action on the complex coordinates
$z_i$: $gz_1=z_1$, $gz_2=\omega z_2$, $gz_3=\omega^{-1} z_3$, $Rz_1=-z_1$, $Rz_2=-z_2$,
$Rz_3=z_3$ (here $\omega=\exp(2\pi i/3)$). This orientifold is ``T-dual'' (in the non-compact limit) of a model studied in \cite{Zwart}. The untwisted tadpole cancellation conditions require presence of one set of D7-branes (whose locations are given by points in the $z_3$ complex plane). The twisted tadpole cancellation conditions read
\cite{KS2,Zwart,AFIV} ($k=1,2$):
\begin{eqnarray}
 &&{\mbox{Tr}}(\gamma_{R,3})={\mbox{Tr}}(\gamma_{R,7})=0~,\\
 &&{\mbox{Tr}}(\gamma_{g^kR,3})={\mbox{Tr}}(\gamma_{g^kR,7})=0~\\
 &&{\mbox{Tr}}(\gamma_{g^k,3})={\mbox{Tr}}(\gamma_{g^k,7})=2~.
\end{eqnarray} 
Note that a choice\footnote{This choice is unique (up to equivalent representations). In particular, any other choice would lead to inconsistent massless spectrum \cite{PS,GP}.} consistent with the six dimensional 
${\bf Z}_2$ model of \cite{PS,GP} (note that ${\bf Z}_2\subset {\bf Z}_6^\prime$ acts as in the
model of \cite{PS,GP}) is given by ($N=n_3/2$)
\begin{eqnarray}
 &&\gamma_{R,3}={\mbox{diag}}(i,-i)\otimes {\bf I}_{N}~,\\
 &&\gamma_{R,7}={\mbox{diag}}(i,-i)\otimes {\bf I}_{4}~.
\end{eqnarray}
However, then the twisted tadpole cancellation conditions for 
$\gamma_{g^k,7}$ and $\gamma_{g^kR,7}$ 
($k=1,2$) have no solution. This is due to the fact that for this orbifold group
we expect non-perturbative contributions from the sectors corresponding to the elements
$\Omega JRg^k$ (as well as the $\Omega Jg^k$ elements - see \cite{KST} for details) of the orientifold group.
These non-perturbative states are present even {\em after} the appropriate blow-ups of the
corresponding orbifold singularities (note that these blow-ups are necessary for consistency
of the orientifold \cite{KST}) are performed. 

{}The above phenomenon occurs for all the other (Abelian) choices of $\Gamma$ which differ from those discussed in the previous section. In particular, let us consider the cases where the
orbifold group is such that it acts crystallographically on $T^6$. For such orbifolds we can
attempt to consider $\Omega$ orientifolds of Type IIB on $T^6/\Gamma$. In these models we have D9-branes, and also D5-branes if $\exists {\bf Z}_2\subset\Gamma$. Let us consider
cases with D5-branes. Then the choices of $\Gamma$ that act crystallographically on $T^6$
are the following: ${\bf Z}_2\otimes {\bf Z}_2$, ${\bf Z}_2\otimes {\bf Z}_2\otimes {\bf Z}_3$, ${\bf Z}_6$ (these are the cases discussed in the previous section), plus
${\bf Z}_6^\prime$, ${\bf Z}_2\otimes {\bf Z}_6$, 
${\bf Z}_3\otimes {\bf Z}_6$, ${\bf Z}_6\otimes {\bf Z}_6$, ${\bf Z}_2\otimes {\bf Z}_4$,
${\bf Z}_4\otimes {\bf Z}_4$, ${\bf Z}_4$,
${\bf Z}_8$, ${\bf Z}_8^\prime$,  ${\bf Z}_{12}$, ${\bf Z}_{12}^\prime$
(see \cite{KST} for details). In the first three cases
the ``naive'' tadpole cancellation conditions give the correct spectra 
for both $\Omega J$ orientifolds of Type IIB on ${\bf C}^3/\Gamma$ and
$\Omega$ orientifolds of Type IIB on $T^6/\Gamma$
(corresponding to the orientifolds after the appropriate blow-ups). In the other cases
the ``naive'' tadpole cancellation conditions have no solution for the 
$\Omega J$ orientifolds of Type IIB on ${\bf C}^3/\Gamma$. As to the $\Omega$ 
orientifolds of Type IIB on $T^6/\Gamma$, the ``naive'' tadpole cancellation conditions
have no solution in the ${\bf Z}_2\otimes {\bf Z}_4$,
${\bf Z}_4\otimes {\bf Z}_4$ \cite{Zwart}, and ${\bf Z}_4$,
${\bf Z}_8$, ${\bf Z}_8^\prime$, ${\bf Z}_{12}^\prime$ \cite{AFIV} cases. In other cases the
``naive'' tadpole cancellation conditions do have solutions and give rise to massless spectra
free of non-Abelian gauge anomalies. However, as explained in \cite{KST}, even in such cases
the massless spectra derived using the world-sheet approach are incomplete. The fact that the
``naive'' tadpoles (and, subsequently, the non-Abelian gauge anomalies in the
``naive'' perturbative spectra) cancel is due to the fact that in these cases the non-perturbative
(from the orientifold viewpoint) states (even though they typically are chiral)
come in combinations such that they do not contribute into the non-Abelian gauge anomaly.
(This was discussed in the ${\bf Z}_6^\prime$ example in \cite{KST}.) On the other hand, the issue of whether such non-perturbative states are indeed present in a given orientifold
is really a {\em local} question as far as the geometry is concerned. In particular, this is due to 
the fact that such non-perturbative states arise in sectors corresponding to certain D-branes
wrapping various (collapsed) 2-cycles at orbifold singularities. Thus, we should be able to
test this issue in a local, that is, non-compact framework. The $\Omega J$ orientifolds of Type IIB on ${\bf C}^3/\Gamma$ provide precisely such a setup. 

{}The above discussion implies that the only (Abelian) orientifolds of Type IIB on $T^6/\Gamma$
with D5-branes for which the ``naive'' (that is, perturbative) description gives the correct tadpoles {\em and} massless spectra (corresponding to orientifolds after the appropriate blow-ups) are
those with $\Gamma\approx{\bf Z}_2\otimes {\bf Z}_2$ \cite{BL}, $\Gamma\approx{\bf Z}_6$ \cite{KS1}, and $\Gamma\approx{\bf Z}_2\otimes {\bf Z}_2\otimes {\bf Z}_3$ \cite{zk}. On the other hand, none of the ${\bf Z}_6^\prime$, ${\bf Z}_2\otimes {\bf Z}_6$, 
${\bf Z}_3\otimes {\bf Z}_6$, ${\bf Z}_6\otimes {\bf Z}_6$
models of \cite{Zwart}, and the ${\bf Z}_{12}$ model of \cite{AFIV} have a perturbative orientifold description. In particular, the ``naive'' perturbative spectra for these models given in \cite{Zwart} and \cite{AFIV} are incomplete. The inadequacy of the world-sheet description is more apparent in the ${\bf Z}_2\otimes {\bf Z}_4$ and
${\bf Z}_4\otimes {\bf Z}_4$ \cite{Zwart}, and ${\bf Z}_4$, ${\bf Z}_8$, ${\bf Z}_8^\prime$ and ${\bf Z}_{12}^\prime$ \cite{AFIV} cases (where the ``naive'' tadpole cancellation conditions have no solution). 

{}We finish this subsection by noting that the crucial check for the consistency of the
${\bf Z}_6$ model of \cite{KS1}, and the ${\bf Z}_2\otimes {\bf Z}_2\otimes {\bf Z}_3$
model of \cite{zk} was utilizing (weak-weak) \cite{ZK,KS1,KS2} Type I-heterotic duality \cite{PW}
in four dimensions. In particular, this duality can be understood in detail \cite{ZK,KS1,KS2}
in the cases without D5-branes to which we turn next.

\subsection{Cases without D7-branes}

{}Let us consider the $\Omega J$ orientifolds of Type IIB on ${\bf C}^3/\Gamma$, where
$\Gamma$ is an Abelian subgroup of $SU(3)$ such that $\not\exists {\bf Z}_2\subset \Gamma$.
Such orientifolds contain only D3-branes but no D7-branes. In the previous section we considered three cases, namely, with $\Gamma\approx{\bf Z}_3$, 
$\Gamma\approx{\bf Z}_7$ and $\Gamma\approx{\bf Z}_3\otimes {\bf Z}_3$. In all these cases
the tadpole cancellation conditions have a solution and yield theories free of non-Abelian gauge anomalies. Note that these are the only Abelian groups without ${\bf Z}_2$ subgroups that act
crystallographically on $T^6$. This implies that the $\Omega$ orientifolds of Type IIB on $T^6/\Gamma$ for these three choices of $\Gamma$ are consistent models (note that these
models contain only D9-branes but no D5-branes) where all the non-perturbative (from the orientifold viewpoint) states should be absent after the appropriate blow-ups. In fact, this was
shown to be the case for the ${\bf Z}_3$ model of \cite{Sagnotti} in \cite{ZK} using the observation \cite{Sagnotti,ZK} that the heterotic dual of this model is perturbative since there are no D5-branes in this orientifold. Subsequently, similar analyses were performed for the
${\bf Z}_7$ \cite{KS1} and ${\bf Z}_3\otimes {\bf Z}_3$ \cite{KS2} cases.

{}Here we would like to address the issue whether for other choices of $\Gamma$ (without ${\bf Z}_2$ subgroups) the $\Omega J$ orientifolds of Type IIB on ${\bf C}^3/\Gamma$ lead to consistent models. The answer to this question is negative. Instead of being most general here,
we will consider two examples. First, consider $\Gamma\approx{\bf Z}_M\otimes {\bf Z}_L$,
where $M,L\in 2{\bf N}+1$, and the generators $g$ and $\theta$ of the ${\bf Z}_M$ and ${\bf Z}_L$ subgroups have the following action on the complex coordinates $z_i$:
$gz_1=\omega z_1$, $gz_2=\omega^{-1} z_2$, $gz_3=z_3$, $\theta z_1=z_1$, $\theta z_2=\eta z_2$, $\theta z_3=\eta^{-1} z_3$ (here $\omega=\exp(2\pi i/M)$, $\eta=\exp(2\pi i/L)$). In
\cite{zura} it was pointed out that the tadpole cancellation conditions for the $\Omega J$ orientifolds of Type IIB on ${\bf C}^3/{\bf Z}_M$ and ${\bf C}^3/{\bf Z}_L$ (where the orbifold groups act as above) have solutions only for $M=L=3$. This ultimately implies that the only
consistent $\Omega J$ orientifold of Type IIB on ${\bf C}^3/ {\bf Z}_M\otimes {\bf Z}_L$ is that
with $M=L=3$, {\em i.e.}, the ${\bf Z}_3\otimes {\bf Z}_3$ model discussed in the previous section.

{}The following example appears to be very illustrative, so we will discuss it in detail. Consider
the orbifold group $\Gamma\approx{\bf Z}_5$ where the generator $g$ of ${\bf Z}_5$ has
the following action on the complex coordinates $z_i$:
$gz_1=\omega z_1$, $gz_2=\omega z_2$, $gz_3=\omega^3 z_3$, where $\omega=
\exp(2\pi i/5)$. Before we discuss the tadpole cancellation conditions in this case (which
follow from the general formula (\ref{tadpoles1})), let us make the most general (up to equivalent
representations) choice for the Chan-Paton matrix $\gamma_{1,3}$ and compute the open string spectrum as we would normally do in the perturbative orientifold approach. Thus, let
($M+2L+2K=n_3$)
\begin{equation}\label{gb}
 \gamma_{1,3}={\mbox{diag}}({\bf I}_M,\omega {\bf I}_L,\omega^{-1} {\bf I}_L, 
 \omega^2 {\bf I}_K,\omega^{-2} {\bf I}_K )~.
\end{equation}
The gauge group is $SO(M)\otimes U(L)\otimes U(K)$ (for definiteness here we have chosen 
the $\Omega$ projection to be of the $SO$ type). The matter charged matter consists of chiral
multiplets in the following irreps of the gauge group (for simplicity we are suppressing the 
$U(1)$ charges):\\
$2\times [({\bf M},{\bf L},{\bf 1})\oplus ({\bf 1},{\overline {\bf L}},{\bf K})\oplus
({\bf 1},{\bf 1},{\overline{\bf A}})]$;\\
$({\bf M},{\bf 1},{\overline {\bf K}})\oplus ({\bf 1},{\bf L},{\bf K})\oplus
({\bf 1},{\overline{\bf A}},{\bf 1})$.\\
It is not difficult to show that this spectrum is free of non-Abelian ($SU(L)$ and $SU(K)$)
gauge anomalies if and only if 
\begin{equation}\label{ML}
 M=L=K-4~.
\end{equation}

{}Now we will show that the above solution for $M$ and $L$ in terms of $K$ is not compatible
with the tadpole cancellation condition which follows from (\ref{tadpoles1}). The latter implies
that
\begin{equation}
 {\mbox{Tr}}(\gamma_{1,3})=4(1+\omega^2+\omega^{-2})~.
\end{equation}  
The solution to this tadpole cancellation condition reads ($N=(n_3-12)/5$):
\begin{equation}\label{soli}
 \gamma_{1,3}={\mbox{diag}}({\bf I}_{N+4},\omega {\bf I}_N,\omega^{-1} {\bf I}_N, 
 \omega^2 {\bf I}_{N+4},\omega^{-2} {\bf I}_{N+4})~.
\end{equation}
Note that (\ref{ML})) and (\ref{soli}) are incompatible.

{}The explanation to the above fact is the same as in the cases discussed in the previous subsection: the non-perturbative (from the orientifold viewpoint) states (arising in the 
$\Omega J g^k$ ($k=1,2,3,4$) sectors) cannot be ignored in this model (as they do not
decouple after the blow-ups). This can be more explicitly seen using Type I-heterotic duality.

{}First we note that the twisted tadpole cancellation conditions in the $\Omega J$
orientifold of Type IIB on ${\bf C}^3/{\bf Z}_5$ are isomorphic to those in the
$\Omega$ orientifold of Type IIB on ${\bf C}^3/{\bf Z}_5$ (after interchanging the corresponding D3- and D9-brane Chan-Paton matrices). Therefore, we can attempt to understand the non-perturbative sectors in the former orientifold via studying the latter.
The above orbifold group ${\bf Z}_5$ does not act crystallographically on $T^6$, so we cannot
build the corresponding compact model. Nonetheless, the
$\Omega$ orientifold of Type IIB on ${\bf C}^3/{\bf Z}_5$ (after the appropriate blow-ups) corresponds to Type I on ${\bf C}^3/{\bf Z}_5$. Its expected heterotic dual is a $Spin(32)/{\bf Z}_2$ heterotic ``compactification'' on ${\bf C}^3/{\bf Z}_5$. We can, therefore, study this model for the above
purposes.

{}Let us choose the $Spin(32)/{\bf Z}_2$ gauge bundle to be the same as
in (\ref{gb}). That is, the action of the orbifold group element $g$ 
on the $Spin(32)/{\bf Z}_2$ lattice is
given by the $16\times 16$ matrix $W$ (in the $SO(32)$ basis) such that $\gamma_{1,3}=W\otimes {\bf I}_2$. The untwisted sector gauge group and charged matter
content then are the same as in the above orientifold model. Note that the equation
$M+2L+2K=32$ subject to the conditions (\ref{ML}) does not have an integer solution. This
implies that the untwisted sector states cannot be free of non-Abelian gauge anomalies. Thus,
in the heterotic model (which we assume to be perturbative subject to the level-matching
constraint which can be satisfied by appropriately choosing the gauge bundle) the twisted sector
states are chiral and have non-trivial contribution to the non-Abelian gauge anomaly. This implies that no blow-ups can get rid of these twisted states (and leave only the untwisted sector
states). The twisted sector states on the heterotic side correspond to the non-perturbative (from the orientifold viewpoint) states on the Type I side. We therefore conclude that these non-perturbative states do not decouple from the massless spectrum (even after the appropriate
blow-ups are performed). That is, the superpotential in the twisted sectors that couples the orbifold blow-up modes and the charged (under the non-Abelian gauge group) twisted states is such that there always are left-over twisted states even for completely blown up orbifold. This is enough to conclude that the non-perturbative (from the orientifold viewpoint) states on the Type I side are important and cannot be ignored (as in the perturbative calculation). 

{}Here one remark is in order. Note that since both Type I and heterotic models are non-compact
(that is, these are ten dimensional theories in the orbifold background that breaks Lorentz invariance) one of the theories is strongly coupled. However, here we are interested in a
general statement about the presence of {\em chiral} states. It is therefore reasonable to assume that the untwisted heterotic spectrum (even at strong coupling) is anomalous without adding chiral states from the twisted sectors. In the corresponding weakly coupled Type I model the
33 open string sector states (that map to the untwisted states on the heterotic side) have non-Abelian gauge anomaly, so some additional states must be present. These states are
in the heterotic picture identified with the twisted sector states (in the strong coupling regime
after blow-ups).

\section{Summary}\label{sum}

{}Let us summarize some of the main conclusions of the previous discussions.\\
$\bullet$ In this paper we considered $\Omega J$ orientifolds of Type IIB on ${\bf C}^3/\Gamma$, where the orbifold group $\Gamma$ is an Abelian finite discrete subgroup of
$SU(3)$. These orientifolds result in  theories with D3-branes. Four dimensional ${\cal N}=1$ space-time supersymmetric gauge theories live in their world-volumes.\\
$\bullet$ There are six orbifold groups for which such theories have perturbative orientifold
description (after appropriately blowing up the orbifold singularities).
These are the ${\bf Z}_3$, ${\bf Z}_7$, ${\bf Z}_3\otimes 
{\bf Z}_3$, ${\bf Z}_2\otimes {\bf Z}_2$, ${\bf Z}_2\otimes {\bf Z}_2\otimes 
{\bf Z}_3$ and ${\bf Z}_2\otimes {\bf Z}_3$ orbifolds discussed in section \ref{N1}. These theories in the large $N$ limit possess the property that computation of any correlation function is reduced to the corresponding computation in the parent ${\cal N}=4$ supersymmetric {\em oriented} gauge theory before orbifolding and orientifolding.\\
$\bullet$ For other Abelian subgroups of $SU(3)$ the world-sheet description of the orientifold theory is inadequate due to non-perturbative states arising from wrapped D-brane sectors.
Such theories can be considered in the F-theory context. However, {\em a priori} there is no reason to expect that the above nice properties will also be present in such gauge theories.\\
$\bullet$ $\Omega J$ orientifolds of Type IIB on ${\bf C}^3/\Gamma$ which do have a world-sheet description are in one-to-one correspondence with the corresponding consistent (from the world-sheet point of view) compact models, that is, the $\Omega$ orientifolds of Type IIB on $T^6/\Gamma$. This provides a strong check for the latter, as well as for the correctness of the arguments of \cite{KST} concerning perturbative {\em vs.} non-perturbative orientifold models. In fact, the matching of arguments in the local and compact  cases, as well as convergence of various approaches to orientifolds discussed in \cite{KST}, clearly indicates that
the understanding developed in \cite{KST} must be rather complete.\\
$\bullet$ It is also rather satisfying to observe that these string theoretic arguments nicely match
({\em a priori} independent) field theory expectations (as in the ${\bf Z}_5$ case discussed in section \ref{other}).

\acknowledgments

{}This work was supported in part by the grant NSF PHY-96-02074, 
and the DOE 1994 OJI award. I would also like to thank Albert and Ribena Yu for 
financial support.

\begin{table}[t]
\begin{tabular}{|c|c|l|c|}
 Model & Gauge Group & \phantom{Hy} Charged  & Twisted Sector 
  \\
       &                &Chiral Multiplets & Chiral Multiplets
 \\
\hline
${\bf Z}_3$ & $U(N)\otimes G_\eta(N+4\eta)$  & 
 $3\times ({\bf R}_\eta,{\bf 1})(+2)$ & $1$
 \\
                &  & $3\times ({\overline {\bf N}},{\bf N+4\eta})(-1)$ &  \\
\hline
${\bf Z}_7$ & $U(N)\otimes U(N)\otimes U(N)\otimes$  & 
 $({\bf R}_\eta,{\bf 1},{\bf 1},{\bf 1})(+2,0,0)$ & $3$
  \\
            &  $G_\eta(N-4\eta)$ & $({\bf N},{\bf 1},{\bf 1},{\bf N-4\eta})(+1,0,0)$ &  \\
               &  & $({\overline {\bf N}},{\bf N},{\bf 1},{\bf 1})(-1,+1,0)$ &  \\
              &  & $({\overline {\bf N}},{\overline {\bf N}},{\bf 1},{\bf 1})(-1,-1,0)$ &  \\
 & & plus cyclic permutations of the& \\
 & & $U(N)\otimes U(N)\otimes U(N)$ irreps &\\  
\hline
${\bf Z}_3\otimes {\bf Z}_3$ & $U(N)\otimes U(N)\otimes U(N)\otimes$  & 
 $({\bf R}_\eta,{\bf 1},{\bf 1},{\bf 1},{\bf 1})(+2,0,0,0)$ & $7$
 \\
                & $U(N+2\eta)\otimes 
 G_\eta(N)$ & $({\overline {\bf N}},{\bf 1},{\bf 1},{\bf 1},{\bf N})(-1,0,0,0)$ &  \\
 & & $({\bf 1},{\overline {\bf N}},{\bf 1},{\bf N+2\eta},{\bf 1})(0,-1,0,+1)$ &\\
 & & $({\bf 1},{\bf 1},{\overline {\bf N}},{\overline {\bf N+2\eta}},{\bf 1})(0,0,-1,-1)$ &\\
 & & $({\bf 1},{\bf N},{\bf N},{\bf 1},{\bf 1})(0,+1,+1,0)$ &\\
 & & plus cyclic permutations of the& \\
 & & $U(N)\otimes U(N)\otimes U(N)$ irreps &\\
\hline
\end{tabular}
\caption{The massless spectra of the ${\cal N}=1$ orientifolds of Type IIB on ${\bf C}^3/
{\bf Z}_3$, ${\bf C}^3/ {\bf Z}_7$ and ${\bf C}^3/{\bf Z}_3\otimes {\bf Z}_3$.
The $U(1)$ charges of the states in the 33 open string sector are
given in parentheses. Also, $G_\eta=SO$ for $\eta=-1$ and $G_\eta=Sp$ for $\eta=+1$
(here we are using the convention that $Sp(2m)$ has rank $m$), 
$R_\eta={\bf A}$ (two-index $N(N-1)/2$ dimensional antisymmetric representation of $U(N)$)
for $\eta=-1$,
and $R_\eta={\bf S}$ (two-index $N(N+1)/2$ dimensional symmetric representation of $U(N)$)
for $\eta=+1$.
By twisted sector chiral multiplets we mean those in the twisted {\em closed} 
string sectors.
The untwisted closed string sector states are not shown.}
\label{Z3} 
\end{table}

\begin{table}[t]
\begin{tabular}{|c|c|l|c|}
 Model & Gauge Group & \phantom{Hy} Charged  & Twisted Sector 
  \\
       &                &Chiral Multiplets & Chiral Multiplets
 \\
\hline
${\bf Z}_2\otimes{\bf Z}_2$ & $Sp(N)_{33} \otimes$  & 
 $3 \times ({\bf A})_{33}$ & $3$
 \\
               &  $\bigotimes_{i=1}^3 Sp(4)_{7_i 7_i}$   
                  & $3 \times ({\bf 6}_i)_{7_i 7_i}$ &  \\
                   &  & $({\bf N};{\bf 4}_i)_{3 7_i}$ &  \\
                &  & $({\bf 4}_i;{\bf 4}_j)_{7_i 7_j}$ &  \\
\hline
${\bf Z}_2\otimes{\bf Z}_2\otimes {\bf Z}_3$ & $[U(N)\otimes Sp(N-2)]_{33} \otimes$  & 
 $3 \times ({\bf A},{\bf 1})(+2)_{33}$ & $7$
 \\
               &  $\bigotimes_{i=1}^3 U(2)_{7_i 7_i}$   
                  & $3 \times ({\overline {\bf N}},{\bf N-2})(-1)_{33}$ &  \\
           &&       $3 \times ({\bf 1}_i)(+2_i)_{7_i7_i}$ &\\
     & & $({\bf N},{\bf 1};{\bf 2}_i)(+1;+1_i)_{3 7_i}$ &  \\
                 & & $({\bf 1},{\bf N-2};{\bf 2}_i)(0;-1_i)_{3 7_i}$ &  \\
                &  & $({\bf 2}_i;{\bf 2}_j)(+1_i;+1_j)_{7_i 7_j}$ &  \\
\hline
${\bf Z}_2\otimes{\bf Z}_3$ & $[U(N)\otimes U(N)\otimes U(N-2)]_{33}$  & 
 $2 \times ({\bf A},{\bf 1},{\bf 1})(+2,0,0)_{33}$ & $3$
 \\
              &  $ \otimes[U(2)\otimes U(2)]_{77}$
                      & $2 \times ({\bf 1},{\overline {\bf A}},{\bf 1})(0,-2,0)_{33}$ &  \\
              &    & $2 \times ({\overline {\bf N}},{\bf 1},{\overline {\bf N-2}})(-1,0,-1)_{33}$ &  \\
              &    & $2 \times ({\bf 1},{\bf N},{\bf N-2})(0,+1,+1)_{33}$ &  \\
              &    & $ ({\bf N},{\overline {\bf N}},{\bf 1})(+1,-1,0)_{33}$ &  \\
              &    & $ ({\overline {\bf N}},{\bf 1},{\bf N-2})(-1,0,+1)_{33}$ &  \\
              &    & $ ({\bf 1},{\bf N},{\overline {\bf N-2}})(0+1,-1)_{33}$ &  \\
               &   & $2 \times ({\bf 1},{\bf 1})(+2,0)_{77}$ &  \\
              &    & $2 \times ({\bf 1},{\bf 1})(0,-2)_{77}$ &  \\
              &    & $ ({\bf 2},{\bf 2})(+1,-1)_{77}$ &  \\
              &    & $ ({\bf N}, {\bf 1},{\bf 1};{\bf 2},{\bf 1})(+1,0,0;+1,0)_{37}$ &  \\
              &    & $ ({\bf 1},{\bf 1},{\bf N-2};{\bf 1},{\bf 2})(0,0,+1;0,+1)_{37}$ &  \\
              &    & $ ({\bf 1},{\overline {\bf N}},{\bf 1};{\bf 1},{\bf 2})(0,-1,0;0,-1)_{37}$ &  \\
              &    & $ ({\bf 1},{\bf 1},{\overline {\bf N-2}};{\bf 2},{\bf 1})(0,0,-1;-1,0)_{37}$ &  \\
\hline
\end{tabular}
\caption{The massless spectrum of the ${\cal N}=1$ orientifold of Type IIB on ${\bf C}^3/
({\bf Z}_2\otimes {\bf Z}_2)$. The semi-colon in the column ``Charged Hypermultiplets'' separates $33$ and $7_i 7_i$ representations.
The notation ${\bf A}$ stands for the two-index antisymmetric (reducible)
representation of $Sp(N)$ in the ${\bf Z}_2\otimes {\bf Z}_2$ model, and for the
two-index antisymmetric representation of $SU(N)$ in the ${\bf Z}_2\otimes
{\bf Z}_2\otimes{\bf Z}_3$ and ${\bf Z}_2\otimes{\bf Z}_3$ models.  
The untwisted closed string sector states are not shown.}
\label{Z2} 
\end{table}

\end{document}